\RequirePackage{amsmath}
\documentclass[twocolumn]{svjour3}

\usepackage{hyperref}
\usepackage{graphicx}
\usepackage{amssymb}
\usepackage{color}
\usepackage{pgfplots}
\usepackage{subcaption}
\usepackage{paralist}
\usepackage[round]{natbib}   

\newcommand{\matr}[1]{\mathbf{#1}}
\newcommand{\Qmat}{\matr{Q}}
\newcommand{\QDmat}{\matr{Q}_\Delta}
\newcommand{\tildeQDmat}{\matr{\tilde Q}_\Delta}

\newcommand{\vect}[1]{\boldsymbol{#1}}
\newcommand{\dt}{\Delta t}

\begin{document}

\title{PFASST-ER: Combining the Parallel Full Approximation Scheme in Space and Time with parallelization across the method}

\author{Ruth Sch\"obel \and Robert Speck }

\institute{
	          Ruth Sch\"obel \at
              Institut f\"ur Numerische Mathematik\\ TU Dresden, Germany \\         
              Tel.: +49 351 463-35546 \\
              \email{ruth.schoebel@tu-dresden.de}               
	      \and
              Robert Speck \at
              J\"ulich Supercomputing Centre\\ Forschungszentrum J\"ulich GmbH, Germany\\         
              Tel.: +49 2461 61-1644 \\
              \email{r.speck@fz-juelich.de}                
}

\date{Received: date / Accepted: date}

\maketitle

\begin{abstract}
To extend prevailing scaling limits when solving time-dependent partial differential equations, the parallel full approximation scheme in space and time (PFASST) has been shown to be a  promising parallel-in-time integrator. 
Similar to a space-time multigrid, PFASST is able to compute multiple time-steps simultaneously and is therefore in particular suitable for large-scale applications on high performance computing systems. 
In this work we couple PFASST with a parallel spectral deferred correction (SDC) method, forming an unprecedented doubly time-parallel integrator.
While PFASST provides global, large-scale ``parallelization across the step'', the inner parallel SDC method allows to integrate each individual time-step ``parallel across the method'' using a diagonalized local Quasi-Newton solver.
This new method, which we call ``PFASST with Enhanced concuRrency'' (PFASST-ER), therefore exposes even more temporal parallelism.
For two challenging nonlinear reaction-diffusion problems, we show that PFASST-ER works more efficiently than the classical variants of PFASST and can be used to run parallel-in-time  beyond the number of time-steps.

\keywords{parallel-in-time integration \and parallel full approximation scheme in space and time \and spectral deferred corrections \and parallelization across the method \and parallelization across the step \and Quasi-Newton}
\end{abstract}

\section{Introduction}

The efficient use of modern high performance computing systems for solving space-time-dependent differential equations has become one of the key challenges in computational science. 
Exploiting the exponentially growing number of processors using traditional techniques for spatial parallelism becomes problematic when, for example, for a fixed problem size communication costs starts to dominate.
Parallel-in-time integration methods have recently been shown to provide a promising way to extend these scaling limits.

As one example, the ``Parallel Full Approximation Scheme in Space and Time'' (PFASST) by Emmett and Minion~\citep{EmmettMinion2012} allows to integrate multiple time-steps simultaneously by using inner iteration of spectral deferred corrections (SDC) on a space-time hierarchy.
It works on the so called composite collocation problem, where each time-step includes a further discretization through quadrature nodes.
This ``parallelization across the steps'' approach~\citep{Burrage1997} targets large-scale parallelization on top of saturated spatial parallelization of partial differential equations (PDEs), where parallelization in the temporal domain acts as a multiplier for standard parallelization techniques in space.
In contrast, ``parallelization across the method'' approaches~\citep{Burrage1997} try to parallelize the integration within an individual time-step.
While this typically results in smaller-scale parallelization in the time-domain, parallel efficiency and applicability of these methods are often more favorable.
Most notably, the ``revisionist integral deferred correction me\-thod'' (RIDC) by Christlieb et al.~\citep{ChristliebEtAl2010} makes use of integral deferred corrections (which are indeed closely related to SDC) in order to compute multiple iterations in a pipelined way.
In~\citep{Speck2018}, different approaches for parallelizing SDC across the method have been discussed, allowing the simultaneous computation of updates on multiple quadrature nodes. 
A much more structured and complete overview of parallel-in-time integration approaches can be found in~\citep{Gander2015_Review}. 
In addition, the Parallel-in-Time website (\url{https://parallel-in-time.org}) offers a comprehensive list of references. 

The key goal of parallel-in-time integrators is to expose additional parallelism in the temporal domain in the cases where classical strategies like parallelism in space are either already saturated or not even possible.
In \citep{ClarkeEtAl2019} the classical Parareal method~\citep{LionsEtAl2001} is used to overcome the scaling limit of a space-parallel simulation of a kinematic dynamo on up to 1600 cores.
The multigrid extension of Parareal, the ``multigrid reduction in time'' method (MGRIT), has been shown to provide significant speedup beyond spatial parallelization~\citep{falgout2017multigrid} for a multitude of problems.
Using PFASST, a space-parallel N-body solver has been extended in~\citep{speck_massively_2012} to run on up to 262\,244 cores, while in~\citep{RuprechtEtAl2013_SC} it has been coupled to a space-parallel multigrid solver on up to 458\,752 cores.

So far, parallel-in-time methods have been implemented and tested either without any additional parallelization techniques or in combination with spatial parallelism.
The goal for this work is to couple two different parallel-in-time strategies in order to extend the overall temporal parallelism exposed by the resulting integrator.
To this end, we take the diagonalization idea for SDC presented in~\citep{Speck2018} (parallel across the method) and use it within PFASST (parallel across the steps).
This way we create an algorithm that computes approximations for different time-steps simultaneously but also works in parallel on each time-step itself.   
Doing so we combine the advantages of both parallelization techniques and create the ``Parallel Full Approximation Scheme in Space and Time with Enhanced concuRrency'' (PFASST-ER), an unprecedented doubly time-parallel integrator for PDEs.
In the next section we will first introduce SDC and PFASST from an algebraic point of view, following~\citep{doi:10.1002/nla.2110,BoltenEtAl2018}.
We particularly focus on nonlinear problems and briefly explain the application of a Newton solver within PFASST.
Then, this Newton solver is modified in Section 3 so that by using a diagonalization approach the resulting Quasi-Newton method can be computed in parallel across the quadrature nodes of each time-step.
In Section 4, we compare different variants of this idea to the classical PFASST implementation along the lines of two nonlinear reaction-diffusion equations.
We show parallel runtimes for different setups and evaluate the impact of the various Newton and diagonalization strategies.
Section 5 concludes this work with a short summary and an outlook.

\section{Parallelization across the steps with PFASST}

We focus on an initial value problem
\begin{align}\label{eq:ODE}
  u_t = f(u),\quad u(0) = u_0
\end{align}
with $u(t), u_0, f(u) \in\mathbb{R}$.
In order to keep the notation simple, we do not consider systems of initial value problems for now, where $u(t) \in\mathbb{R}^N$.
Necessary modifications will be mentioned where needed.
In a first step, we now discretize this problem in time and review the idea of single-step, time-serial spectral deferred corrections (SDC).

\subsection{Spectral deferred corrections}

For one time-step on the interval $[t_l,t_{l+1}]$ the Picard formulation of Equation \eqref{eq:ODE} is given by
\begin{align}
  u(t) = u_{l,0} + \int_{t_0}^t f(u(s))ds,\ t\in[t_l,t_{l+1}].
\end{align}

To approximate the integral we use a spectral quadrature rule.
We define $M$ quadrature nodes $\tau_{l,1},...,\tau_{l,M}$, which are given by $t_l \le \tau_{l,1} < ... < \tau_{l,M} = t_{l+1}$. 
We will in the following explicitly exploit the condition that the last node is equal to the right integral boundary. 
Quadrature rules like Gau\ss-Radau or Gau\ss-Lobatto quadrature satisfy this property. 
We can then approximate the integrals from $t_l$ to the nodes $\tau_{l,m}$, such that
\begin{align*}
  u_{l,m} = u_{l,0} + \Delta t \sum_{j=1}^Mq_{m,j}f(u_{l,j}),
\end{align*} 
where $u_{l,m} \approx u(\tau_{l,m})$, $\dt = t_{l+1}-t_l$ and $q_{m,j}$ represent the quadrature weights for the interval $[t_l,\tau_{l,m}]$ such that
\begin{align*}
  \sum_{j=1}^Mq_{m,j}f(u_{l,j})\approx\int_{t_l}^{\tau_{l,m}}f(u(s))ds.
\end{align*}
We combine these $M$ equations into one system
\begin{align}\label{eq:coll_prob}
  \left(\matr{I} - \dt\Qmat\vect{f} \right)(\vect{u}_l) = \vect{u}_{l,0},
\end{align}
which we call the ``collocation problem''.
Here, $\vect{u}_l = (u_{l,1}, ..., u_{l,M})^T \approx (u(\tau_{l,1}), ..., u(\tau_{l,M}))^T\in\mathbb{R}^M$, $\vect{u}_{l,0} = (u_{l,0}, ..., u_{l,0})^T\in\mathbb{R}^M$, $\Qmat = (q_{ij})_{i,j}\in\mathbb{R}^{M\times M}$ 
is the matrix gathering the quadrature weights and the vector function $\vect{f}:\mathbb{R}^M \to \mathbb{R}^M$ is given by 
\begin{align*}
    \vect{f}(\vect{u}_l) = (f(u_{l,1}), ..., f(u_{l,M}))^T.
\end{align*}
To simplify the notation we define
\begin{align*}
  \matr{C}^{\operatorname{coll}}_{\vect f}(\vect{u}_l) := \left(\matr{I} - \dt\Qmat\vect{f} \right)(\vect{u}_l)  .
\end{align*}
We note that for $u(t) \in\mathbb{R}^N$, we need to replace $\Qmat$ by $\Qmat\otimes\matr{I}_N$.

\bigskip

System \eqref{eq:coll_prob} is dense and a direct solution is not advisable, in particular if $\vect{f}$ is a nonlinear operator.
The spectral deferred correction method solves the collocation problem in an iterative way. 
While it has been derived originally from classical deferred or defect correction strategies, we here follow \citep{HuangEtAl2006,Weiser2014,RuprechtSpeck2016} to present SDC as
preconditioned Picard iteration. 
A standard Picard iteration is given by
\begin{align*}
  \vect{u}^{k+1}_l = \vect{u}^{k}_l + (\vect{u}_{l,0} -  \matr{C}^{\operatorname{coll}}_{\vect f} (\vect{u}^k_l))
\end{align*} 
for $k = 0, \dots, K$, and some initial guess $\vect{u}^{0}_l$.

In order to increase range and speed of convergence, we now precondition this iteration.
The standard approach to preconditioning is to define an operator $\matr{P}^{\operatorname{sdc}}_{\vect f}$, which is easy to invert but also close to the operator of the system. 
We define this ``SDC preconditioner'' as
\begin{align*}
  \matr{P}^{\operatorname{sdc}}_{\vect f}(\vect{u}_l) := \left(\matr{I} - \dt\QDmat \vect{f} \right)(\vect{u}_l)  
\end{align*}
so that the preconditioned Picard iteration reads
\begin{align} \label{eq:SDC}
 \matr{P}^{\operatorname{sdc}}_{\vect f}(\vect{u}_l^{k+1}) =  (\matr{P}^{\operatorname{sdc}}_{\vect f} -  \matr{C}^{\operatorname{coll}}_{\vect f})(\vect{u}_l^k) + \vect{u}_{l,0}  .
\end{align} 
The key for defining $\matr{P}^{\operatorname{sdc}}_{\vect f}$ is the choice of the matrix $\QDmat$.
The idea is to choose a ``simpler'' quadrature rule to generate a triangular matrix $\QDmat$ such that solving System \eqref{eq:SDC} can be done by forward substitution.
Common choices include the implicit Euler method or the so-called ``LU-trick'', where the LU decomposition of $\Qmat^T$ with 
\begin{align}\label{eq:lu}
  \QDmat^{\mathrm{LU}} = \matr{U^T}\quad \text{for}\quad \Qmat^T = \matr{L}\matr{U}
\end{align}
is used~\citep{Weiser2014}.

System~\eqref{eq:SDC} establishes the method of spectral deferred corrections, which can be used to approximate the solution of the collocation problem on a single time-step.
In the next step, we will couple multiple collocation problems and use SDC to explain the idea of the parallel full approximation scheme in space and time.

\subsection{Parallel full approximation scheme in space and time}

The idea of PFASST is to solve a ``composite collocation problem'' for multiple time-steps at once using multigrid techniques and SDC for each step in parallel.
This composite collocation problem for $L$ time-steps can be written as
 \begin{align*} 
   \begin{pmatrix}
    \matr{C}^{\operatorname{coll}}_{\vect f} \\
     -\matr{H} & \matr{C}^{\operatorname{coll}}_{\vect f} \\
      & \ddots & \ddots \\
     & & -\matr{H} & \matr{C}^{\operatorname{coll}}_{\vect f}
   \end{pmatrix}
   \begin{pmatrix}
     \vect{u}_{1}\\
     \vect{u}_{2}\\
     \vdots\\
     \vect{u}_{L}
   \end{pmatrix} = 
   \begin{pmatrix}
     \vect{u}_{0,0} \\
     \vect 0\\
     \vdots\\
     \vect 0
   \end{pmatrix},
 \end{align*}
where the matrix $\matr{H}\in\mathbb{R}^{M\times M}$ on the lower subdiagonal transfers the information from one time-step to the next one.
It takes the value of the last node $\tau_{l,M}$ of an interval $[t_l, t_{l+1}]$, which is by requirement equal to the left boundary $t_{l+1}$ of the following interval $[t_{l+1}, t_{l+2}]$, and provides it as a new starting value for this interval. 
Therefore, the matrix $\matr{H}$ contains the value $1$ on every position in the last column and zeros elsewhere.
To write the composite collocation problem in a more compact form we define the vector $\vect{u} = (\vect{u}_{1}, ..., \vect{u}_{L})^T\in\mathbb{R}^{LM}$, which contains the solution at all quadrature nodes at all time-steps, 
and the vector $\vect{b} = (\vect{u}_{0,0}, \vect 0, ..., \vect 0)^T\in\mathbb{R}^{LM}$, which contains the initial condition for all nodes at the first interval and zeros elsewhere.
We define ${\vect F: \mathbb{R}^{LM} \rightarrow \mathbb{R}^{LM}}$ as an extension of $\vect f$ so that
${\vect F} ({\vect u}) = \left( {\vect f} ({\vect u}_{1}), \dots , {\vect f} ({\vect u}_{L})  \right)^T$.
Then, the composite collocation problem can be written as 
\begin{align}
    \matr C_{\vect F}(\vect u) = \vect b. \label{eq:comp_coll_prob}
\end{align}
with
\begin{align*}
    \matr C_{\vect F}(\vect u) = \left(\matr{I}  - \dt (\matr{I}_L\otimes\matr{Q} )\vect F - \matr{E}\otimes\matr{H}\right)(\vect u),
\end{align*}
where the matrix $\matr{E}\in\mathbb{R}^{L\times L}$ just has ones on the first subdiagonal and zeros elsewhere.
If $u \in\mathbb{R}^N$, we need to replace $\matr{H}$ by $\matr{H}\otimes\matr{I}_N$. 

\bigskip

SDC can be used to solve the composite collocation problem by forward substitution in a sequential way.
As a parallel-in-time integrator PFASST is an attractive alternative.
The first step from SDC towards PFASST is the introduction of multiple levels, which are representations of the problem with different accuracies in space and time. 
In order to simplify the notation we focus to a two-level scheme consisting of a fine and a coarse level.
Coarsening can be achieved for example by reducing the resolution in space, by decreasing the number of quadrature nodes on each interval or by solving implicit systems less accurately.
For this work, we only consider coarsening in space, i.e., by using a restriction operator ${R}$ on a vector $u\in\mathbb{R}^{N}$ we obtain a new vector $\tilde{u}\in\mathbb{R}^{\tilde{N}}$.
Vice versa, the interpolation operator ${T}$ is used to interpolate values from $\tilde{u}$ to $u$.
Operators, vectors and numbers on the coarse level will be denoted by a tilde to avoid further index cluttering.
Thus, the composite collocation operator on the coarse-level is given by $ \matr{\tilde C}_{\vect F}$.
While $\matr{C}_{\vect{F}}$ is defined on $\mathbb{R}^{LMN}$, $\matr{\tilde C}_{\vect F}$ acts on $\mathbb{R}^{L M \tilde N}$ with $\tilde N \le N$, but as before we will neglect the space dimension in the following notation.
The extension of the spatial transfer operators to the full space-time domain is given by $\matr{R} = \matr{I} \otimes R$ and $\matr{T} = \matr{I} \otimes T$.
    
\bigskip

The main goal of the introduction of a coarse level is to move the serial part of the computation to this hopefully cheap level, while being able to run the expensive part in parallel.
For that, we define two preconditioners: a serial one with a lower subdiagonal for the coarse level and a parallel, block-diagonal one for the fine level.
The serial preconditionier for the coarse level is defined by
  \begin{align*}
    \matr{\tilde P}_{\vect F} = 
    \begin{pmatrix}
      \matr{\tilde P}_{\vect f}^{\operatorname{sdc}} \\
      -\matr{\tilde H} & \matr{\tilde P}_{\vect f}^{\operatorname{sdc}} \\
       & \ddots & \ddots \\
      & & -\matr{\tilde H} & \matr{\tilde P}_{\vect f}^{\operatorname{sdc}} \\
    \end{pmatrix},
  \end{align*}
or, in a more compact way, by
\begin{align*}
 &\matr{\tilde P}_{\vect F}(\vect{\tilde u}) = \left( \matr{\tilde I} - \dt(\matr{I}_L \otimes \tildeQDmat )\matr {\tilde F} - \matr{E}\otimes\matr{\tilde H} \right) (\vect{\tilde u}).
\end{align*}
Inverting this corresponds to a single inner iteration of SDC (a ``sweep'') on step 1, then sending forward the result to step 2, an SDC sweep there and so on.
The parallel preconditioner on the fine level then simply reads
\begin{align*}
 &\matr P_{\vect F}(\vect u) = (\matr{I} - \dt (\matr{I}_L\otimes\QDmat ) \vect F) (\vect u).
\end{align*}
Applying $\matr P_{\vect F}$ on the fine level leads to $L$ decoupled SDC sweeps, which can be run in parallel. 

For PFASST, these two preconditioners and the levels they work on are coupled using a full approximation scheme (FAS) known from nonlinear multigrid theory~\citep{trottenberg}.
Following~\citep{doi:10.1002/nla.2110} one iteration of PFASST can then be formulated in four steps:
  \begin{enumerate}
   \item the computation of the FAS correction ${\tau}^k$, including the restriction of the fine value to the coarse level
   \begin{align*}
      {\tau}^k =\matr{\tilde C}_{\vect F} (\matr R {\vect u}^k) - \matr R \matr{ C}_{\vect F} ( {\vect u}^k) ,
   \end{align*} 
   \item the coarse sweep on the modified composite collocation problem on the coarse level
  \begin{align}\label{eq:coarse_sweep}
     \matr{\tilde P}_{\vect F} (\vect{\tilde{u}}^{k+1} ) &= (\matr{\tilde P}_{\vect F} - \matr{\tilde C}_{\vect F})({\vect{ \tilde{u}}}^{k})  +  \vect{\tilde b} + \tau^k,
  \end{align}
   \item the coarse grid correction applied to the fine level value
   \begin{align}\label{update}
     \vect{u}^{k+\frac{1}{2}} &= \vect{u}^{k} + \matr T ( \vect{\tilde{u}}^{k+1} -\matr R \vect{u}^k ), 
  \end{align}
   \item the fine sweep on the composite collocation problem on the fine level
  \begin{align}\label{eq:fine_sweep}
     \matr{ P}_{\vect F} ( \vect{u}^{k+1} ) &= (\matr{ P}_{\vect F} - \matr{C}_{\vect F})( \vect{u}^{k+\frac{1}{2}} )  + \vect b . 
  \end{align} 
  \end{enumerate}

  \begin{figure}[t]
  	\centering	
  	\includegraphics[scale=0.8]{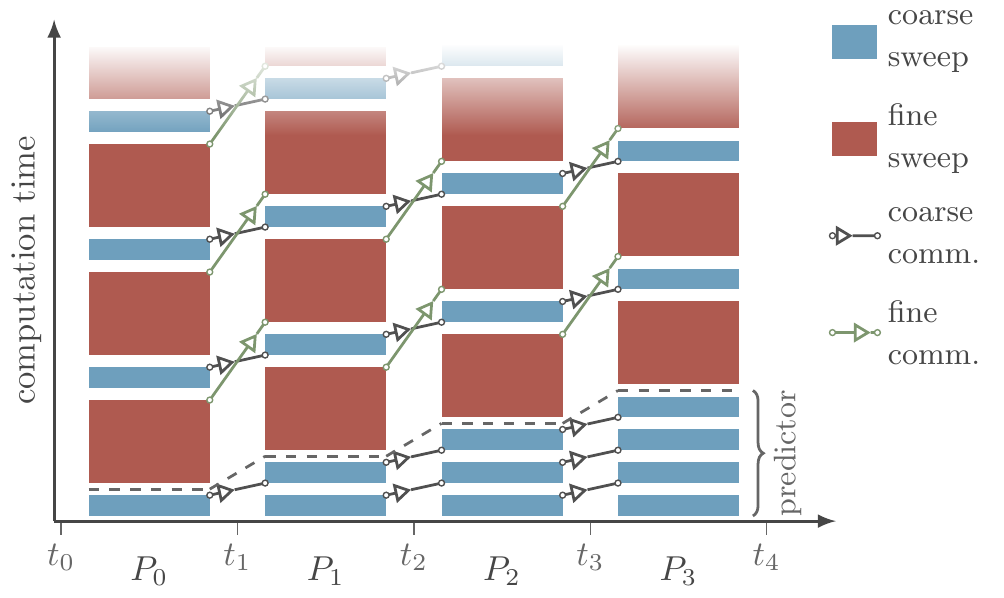}
  	\caption{
  	  Schematic view of PFASST on four processors. 
  	  The figure was created with \texttt{pfasst-tikz}~\citep{pfasst_tikz}.}
  	\label{fig:pfasst_full}
  \end{figure}

 In Figure \ref{fig:pfasst_full}, we see a schematic representation of the described steps. 
 The time-step parallel procedure, which we describe here is also the same for all PFASST versions, that we will introduce later.
 It is common to use as many processors as time-steps: In the given illustration four processors work on four time-steps.
  Therefore the temporal domain is divided into four intervals, which are assigned to four processors $P_0, ..., P_3$.
  Every processor performs SDC sweeps on its assigned interval on alternating levels.
  The big red blocks represent fine sweeps, given by Equation \eqref{eq:fine_sweep}, and the small blue blocks coarse sweeps, given by Equation \eqref{eq:coarse_sweep}.
  The coarse sweep over all intervals is a serial process: after a processor finished its coarse sweeps, it sends forward its results to the next processor, that take this result as an initial value for its own coarse sweeps.
  We see the communication in the picture represented by small arrows, which connect the coarse sweeps of each interval.
  In formula \eqref{eq:coarse_sweep}, the need for communication with a neighboring process is obvious, because $\matr{\tilde P}_{\vect F}$ is not a (block-) diagonal matrix, but has entries on its lower block-diagonal.
  $\matr{P}_{\vect F}$ on the other hand is block-diagonal, which means that the processors can calculate on the fine level in parallel.
  We see in Formula \eqref{eq:fine_sweep} that there is only a connection to previous time-steps through the right-hand side, where we gather values from the previous time-step and iteration but not from the current iteration.
  The picture shows this connection by a fine communication, which forwards data from each fine sweep to the following fine sweep of the right neighbor.
  The fine and coarse calculations on every processor are connected through the FAS corrections, which in our formula are part of the coarse sweep.

\subsection{PFASST-Newton}

For each coarse and each fine sweep within each PFASST iteration, System \eqref{eq:coarse_sweep} and System \eqref{eq:fine_sweep}, respectively, need to be solved.
If $f$ is a nonlinear function these systems are nonlinear as well.
The obvious and traditional way to proceed in this case is to linearize the problem using Newton's method.
This way, PFASST is the outer solver with an inner Newton iteration.
For triangular $\QDmat$, the $m$th equation on the $l$th time-step on the coarse level reads
\begin{align*}
    (1 - \dt\ \tilde q^\Delta_{l,m} \tilde f) (\tilde u^{k+1}_{l,m}) = &\ \tilde u^{k+1}_{l, 0} \\&+ \dt \sum_{n=1}^{m-1} \tilde q^\Delta_{l,n} \tilde f(\tilde u^{k+1}_{l,n}) \\&+ \vect{\tilde c}( \vect{\tilde u}^k)_{l,m},
\end{align*}
where $\tilde u^{k+1}_{0,0} = \tilde u_{0,0}$ and $\vect{\tilde c}(\vect{\tilde u}^k)_{l,m}$ is the $m$th entry the $l$th block of 
$\vect{\tilde c}( \vect{\tilde u}^k) :=  (\matr{\tilde P}_{\vect F} - \matr{\tilde C}_{\vect F})({\vect{ \tilde{u}}}^{k})   + \tau^k.$
This term gathers all values of the previous iteration.
The first summand of the right-hand side of the coarse level equation corresponds to the $\tilde{\vect{b}}$ and the $\tilde{\matr{H}}$, while the following sum comes from the lower triangular structure of $\tildeQDmat$.

For time-step $l$ these equations can be solved one by one using Newton iterations and forward substitution.
This is inherently serial, because the solution on the $m$th quadrature node depends on the solution at all previous nodes through the sum.
Thus, while running parallel across the steps, each of the solution of the local collocation problems is found in serial.
In the next section, we will present a novel way of applying Newton's method, which allows to parallelize this part across the collocation nodes, joining parallelization across the step with parallelization across the method.
We call this method PFASST-ER: the ``Parallel Full Approximation Scheme in Space and Time with Enhanced concuRrency''.

\section{PFASST-ER}

From the perspective of a single time-step $[t_l, t_{l+1}]$ or processor $P_l$, equation~\eqref{eq:coarse_sweep} on the coarse level for this step reads
\begin{align*}
    \matr{\tilde P}_{\vect f}^{\operatorname{sdc}}(\tilde{\vect{u}}_{l}^{k+1}) - \tilde{\vect{u}}_{l,0}^{k+1} =&\ (\matr{\tilde P}_{\vect f}^{\operatorname{sdc}} - \matr{\tilde C}_{\vect f}^{\operatorname{coll}})(\tilde{\vect{u}}_{l}^{k}) +  \tau^k_{l},
\end{align*}
where $\tau^k_{l}$ is the $l$th component of $\tau^k$, belonging to the interval $[t_l, t_{l+1}]$.
Note that the serial dependency is given by the term $\tilde{\vect{u}}_{l,0}^{k+1}$, so that it does not depend on the solution $\tilde{\vect{u}}_{l}^{k+1}$ of this equation and can thus be considered as part of a given right-hand side.
On the fine level, this is even simpler, because there we have to solve
\begin{align*}
    \matr{P}_{\vect f}^{\operatorname{sdc}}({\vect{u}}_{l}^{k+1}) = (\matr{P}_{\vect f}^{\operatorname{sdc}} - \matr{C}_{\vect f}^{\operatorname{coll}})(\vect{u}_{l}^{k+\frac{1}{2}}) + \vect{u}_{l,0}^{k+\frac{1}{2}} ,
\end{align*}
making the $\vect{u}_{l,0}^{k+\frac{1}{2}}$-term not even dependent on the current iteration (which, of course, leads to the parallelism on the fine level).

As we have seen above, the typical strategy would be to solve these systems line by line, node by node, using forward substitution and previous PFASST iterates as initial guesses.
An alternative approach has been presented in~\citep{Speck2018}, where each SDC iteration can be parallelized across the node.
While this is trivial for linear problems, nonlinear ones require the linearization of the full equations, not node-wise as before.
For the fine sweep, let
\begin{align*}
    \matr{G}_{\vect f}^{\operatorname{sdc}}(\vect{v}) :=&\ \matr{P}_{\vect f}^{\operatorname{sdc}}(\vect{v}) - (\matr{P}_{\vect f}^{\operatorname{sdc}} - \matr{C}_{\vect f}^{\operatorname{coll}})(\vect{u}_{l}^{k+\frac{1}{2}})  -\vect{u}_{l,0}^{k+\frac{1}{2}}
\end{align*}
then a Newton step for $\matr{G}_{\vect f}^{\operatorname{sdc}}(\vect{v}) = 0$ is given by
\begin{align*}
    \nabla\matr{G}_{\vect f}^{\operatorname{sdc}}(\vect{v}^j)\vect{e}^{j} &= -\matr{G}_{\vect f}^{\operatorname{sdc}}(\vect{v}^j),\\
    \vect{v}^{j+1} &= \vect{v}^j + \vect{e}^{j},
\end{align*}
for Jacobian matrix $\nabla\matr{G}_{\vect f}^{\operatorname{sdc}}(\vect{v}^j)$ of $\matr{G}_{\vect f}^{\operatorname{sdc}}$ evaluated at $\vect{v}^j$.
We have
\begin{align*}
    \nabla\matr{G}_{\vect f}^{\operatorname{sdc}}(\vect{v}^j) &= \nabla \matr{P}_{\vect f}^{\operatorname{sdc}}(\vect{v}^j) \\
    &=\matr{I} - \dt\QDmat \nabla\vect{f}(\vect{v}^j)
\end{align*}
for Jacobian matrix $\nabla\vect{f}(\vect{v}^j)$ of $\vect f$ evaluated at $\vect{v}^j$ which in turn is given by
\begin{align*}
    \nabla\vect{f}(\vect{v}^j) = \mathrm{diag}(f'(v_1^j), ..., f'(v_M^j))^T.
\end{align*}
There is still no parallelism to exploit, but when we replace the full Jacobian matrix $\nabla\vect{f}(\vect{v}^j)$ by the approximation $f'(v_{l,0})\matr{I}_M$, which is the derivative of $f$ at the initial value for the current time-step, we can use
\begin{align*}
    \nabla\matr{G}_{\vect f}^{\operatorname{sdc}}(\vect{v}^j) \approx \nabla\matr{G}_{\vect f}^{\Delta\text{-}\mathrm{QN}}(v_{l,0}) := \matr{I} - f'(v_{l,0})\dt\QDmat
\end{align*}
to establish a Quasi-Newton iteration as 
\begin{align*}
    \nabla\matr{G}_{\vect f}^{\Delta\text{-}\mathrm{QN}}(v_{l,0})\vect{e}^{j} &= -\matr{G}_{\vect f}^{\operatorname{sdc}}(\vect{v}^j),\\
    \vect{v}^{j+1} &= \vect{v}^j + \vect{e}^{j}.
\end{align*}
This decouples the evaluation of the Jacobian matrix from the current quadrature nodes and now $\QDmat$ can be diagonalized, so that the inversion of $\nabla\matr{G}_{\vect f}^{\Delta\text{-}\mathrm{QN}}(v_{l,0})$ can be parallelized across the nodes.
Note that there are other options for approximating the full Jacobian matrix. 
Most notably, in~\citep{GanderEtAl2016} the mean over all Jacobian matrices is used (there across the time-steps).
We did not see any impact on the convergence when following this strategy, most likely because the number of quadrature nodes is typically rather low.
The advantage of using the initial value is that it reduces the number of evaluations of the Jacobian matrix, which also includes communication time.

With $\QDmat = \matr{V}_\Delta\matr{\Lambda}_\Delta\matr{V}^{-1}_\Delta$ the algorithm reads:
\begin{enumerate}
  \item replace $\vect{r}^j = -\matr{G}_{\vect f}^{\operatorname{sdc}}(\vect{v}^{j})$ by $\bar{\vect{r}}^j = -\matr{V}_\Delta^{-1}\matr{G}_{\vect f}^{\operatorname{sdc}}(\vect{v}^{j})$ (serial),
  \item solve $\left(\matr{I} - f'(v_{l,0})\dt\matr{\Lambda}_\Delta\right)\bar{\vect{e}}^{j} = \bar{\vect{r}}^j$ (parallel in $M$),
  \item replace $\bar{\vect{e}}^{j}$ by $\vect{e}^{j} = \matr{V}_\Delta\bar{\vect{e}}^{j}$ (serial),
  \item set $\vect{v}^{j+1} = \vect{v}^{j} + \vect{e}^{j}$ (parallel in $M$).
\end{enumerate}
This can be iterated until a certain threshold is reached and then set $\vect{u}^{k+1}_{l} = \vect{v}^J$ to obtain the solution of the equation for the fine sweep.
On the coarse level, the procedure is very similar, with a slightly different definition of $\tilde{\matr{G}}_{\tilde{\vect f}}^{\operatorname{sdc}}(\vect{\tilde{v}})$.
In practice, choosing $J=1$ is sufficient, because this is already the inner solver for an outer PFASST iteration.

\bigskip

This linearization and diagonalization strategy immediately suggests a second approach: instead of using $\QDmat$ for the preconditioner, we can use the original quadrature matrix $\Qmat$ directly.
The intention of using $\QDmat$ in the first place was to obtain a preconditioner which allowed inversion using forward substitutions. 
Now, with diagonalization in place, this is no longer necessary. 
Instead, we can use 
\begin{align*}
    \matr{P}_{\vect f}^{\operatorname{coll}} := \matr{C}_{\vect f}^{\operatorname{coll}}
\end{align*}
and thus
\begin{align*}
    \matr{G}_{\vect f}^{\operatorname{coll}}(\vect{v}) := \matr{C}_{\vect f}^{\operatorname{coll}}(\vect{v}) - \vect{u}_{l,0}^{k+\frac{1}{2}}.
\end{align*}
Note that this is just the $l$th block of the original composite collocation problem.
Following the same ideas as before, we end up with
\begin{align*}
    \nabla\matr{G}_{\vect f}^{\operatorname{coll}}(\vect{v}^j) \approx \nabla\matr{G}_{\vect f}^{\mathrm{QN}}(v_{l,0}) := \matr{I} - f'(v_{l,0})\dt\Qmat,
\end{align*}
which can be diagonalized using $\Qmat = \matr{V}\matr{\Lambda}\matr{V}^{-1}$.
The same idea can be applied to the coarse level sweep, of course.
As a result, the original nonlinear SDC sweeps within PFASST are now replaced by Quasi-Newton iterations
which can be done parallel across the nodes.
We refer to~\citep{Speck2018} for more details on the idea of parallel SDC sweeps with $\Qmat$ and $\QDmat$.

The question now is, how much the approximation of the Jacobians affects the convergence and runtime of the method and how all this compares to standard PFASST iterations with nonlinear SDC.
It is well known that for suitable right-hand sides and initial guesses the standard, unmodified Newton method converges quadratically while the Quasi-Newton method as well as SDC show linear convergence, see e.g.~\citep{Kelley95,Jackson_NewtonRK,Tang2013}.
We will examine the impact of these approaches in the following section along the lines of two numerical examples.
A more rigorous mathematical analysis is currently ongoing work, as it can be embedded into a larger convergence theory for PFASST with inner Newton-type solvers.

\section{Numerical Results}

We apply PFASST and PFASST-ER to two different, rather challenging reaction-diffusion problems, starting with a detailed analysis of the parallelization strategies for the Allen-Cahn equation and highlighting differences to these findings for the Gray-Scott equations.

\subsection{Allen-Cahn equation}

We study the two-dimensional Allen-Cahn equation, which is given by
  \begin{align}
   u_t = \Delta u + \frac{1}{\varepsilon^2} u (1 - u)   
  \end{align}
on the spatial domain $[-0.5,0.5]^2$ and with initial condition
  \begin{align*}
  \begin{split}
  u_0 = \text{tanh}\left( \frac{R_0 -(x^2 + y^2 )}{\sqrt{2} \varepsilon}  \right).                     
  \end{split} 
  \end{align*}    
 
 We use simple second-order finite differences for discretization in space and take $256$ elements in each dimension on the fine level and $128$ on the coarse one.
 We furthermore use $M = 4$ Gau\ss-Radau nodes, set $\varepsilon = 0.04$, $\Delta t = 0.001 < \varepsilon^2$  and stop the
 simulation after $24$ time-steps at $T = 0.024$.
 The initial condition describes a circle with a radius $R_0 = 0.25$, see e.g.~\citep{zhang_SISC_AC}.

 The results we present in the following were computed with pySDC~\citep{Speck2019,pySDC-website} on the supercomputer JURECA~\citep{jureca}.
 We run a serial single-level simulation using SDC (``SL'' in the plots), a serial multi-level simulation using multi-level SDC (``ML'', which is PFASST on one processor, see~\citep{mlsdc-2}) and parallel simulations with $2$, $4$, $8$, $12$ and $24$ processors (``P2'' to ``P24''), all until a given residual tolerance of $10^{-10}$ is reached.

 \begin{figure}[t]
\includegraphics[scale=0.9]{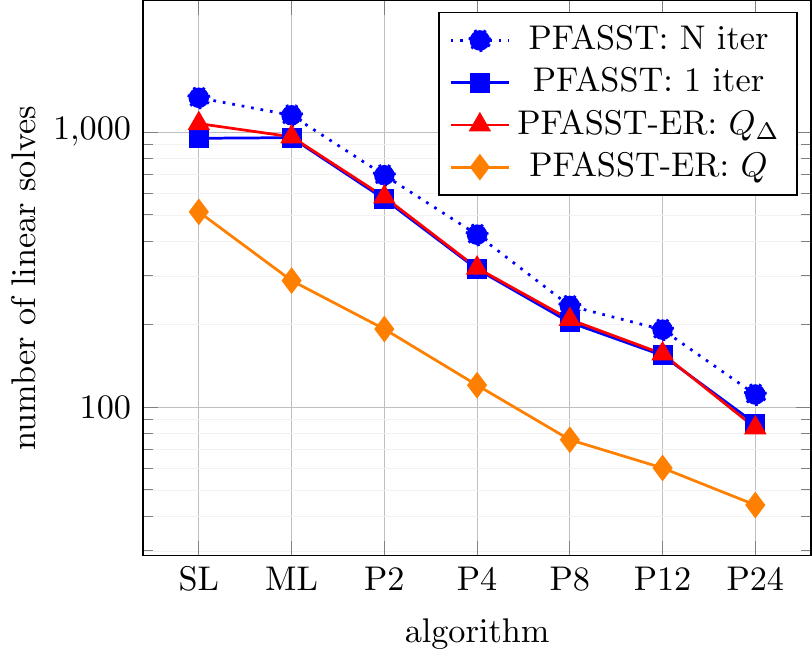}
     \caption{Number of linear solves for the Allen-Cahn example, all methods run serial on the nodes.}
     \label{AC:iterations}
 \end{figure}
 
 In Figure \ref{AC:iterations} we show the number of linear solves for different versions of the solvers, aggregated over all time-steps, quadrature nodes, outer and inner iterations. 
 Here, two versions of the original PFASST algorithm are run: 
 The first one performs exactly one inner Newton iteration in every PFASST iteration; this version is labeled as ``PFASST: 1 iter''. 
 In contrast, ``PFASST: N iter'' performs so many inner Newton iterations that the residual of the nonlinear inner problem is smaller than $10^{-11}$.
 Both PFASST versions use the quadrature matrix $\matr{Q}_\Delta^{LU}$ from Equation~\eqref{eq:lu} inside the preconditioner.
 For PFASST-ER we also differentiate between two variants:
 The PFASST-ER algorithm, which uses the original $\matr Q$ inside the preconditioner is labeled as "PFASST-ER: $Q$" and the one which uses $\matr{Q}^{LU}_\Delta$ is labeled as ``PFASST-ER: $Q_\Delta$''.
 Solving the innermost linear systems is done using GMRES.
 
 We can see that performing more than one inner Newton iteration (``PFASST: N iter'' vs.\ ``PFASST: 1 iter'') does not improve the convergence of the overall algorithm.
 On the contrary more iterations are needed to achieve the same result.
 Using the Quasi-Newton approach with the same preconditioner instead of the classical Newton solver (``PFASST-ER: $Q_\Delta$'' vs.\ ``PFASST: 1 iter'') only shows little effect on the total iteration numbers, but using the original quadrature matrix $\matr Q$ instead of $\matr{Q}^{LU}_\Delta$ inside the preconditioner (``PFASST-ER: $Q$'' vs.\ ``PFASST-ER: $Q_\Delta$'') greatly reduces the number of iterations. 

However, without parallelization one iteration of PFASST-ER with $\matr Q$ is in general more expensive as one iteration of one of the other algorithms, because it requires the solution of a full system via diagonalization instead of stepping through a triangular system via forward substitution.
\begin{figure}[t]
  \includegraphics[scale=0.9]{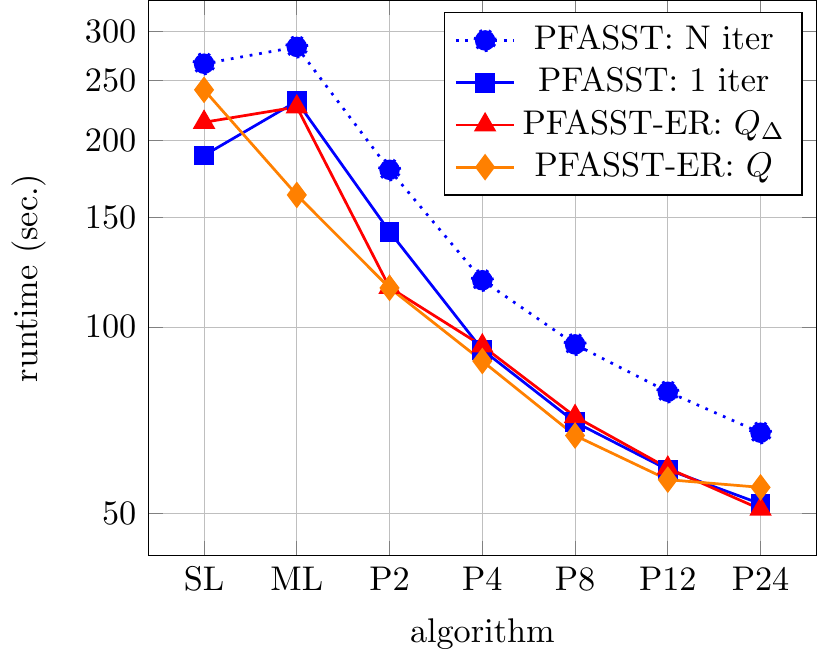} 
  \caption{Time to solution for Allen-Cahn with parallelization only across time-steps.}
  \label{AC:runtime}
\end{figure}
In Figure~\ref{AC:runtime}, we thus examine whether the lower number of higher costly iterations actually pays off.
The plot shows results for the same setup as Figure \ref{AC:iterations}, but now we focus on the runtime instead of the iteration numbers.
We only consider parallelization across the time-steps to compare the impact of the algorithmic change first.
We see that despite the fact that the iterations are much more expensive, PFASST-ER with $\matr Q$ already in this example shows a lower runtime than the original PFASST method.
This is also true when using $\QDmat$ instead of $\matr Q$.
 
 \bigskip

Until now we did not yet consider the additional direction of concurrency exposed by PFASST-ER.
For that, we next compare different distributions of up to $24$ cores on the $4$ quadrature nodes and the $24$ time-steps.
\begin{figure}[t]
    \includegraphics[width=0.425\textwidth]{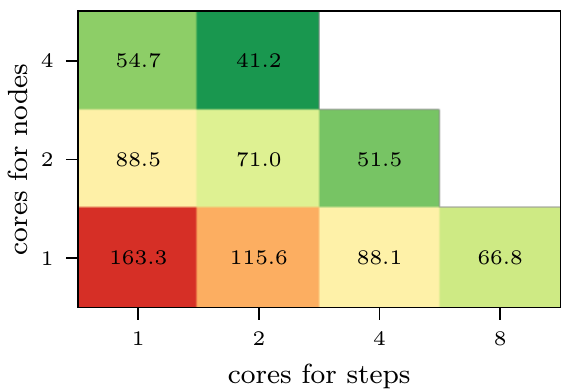}   
    \includegraphics[width=0.425\textwidth]{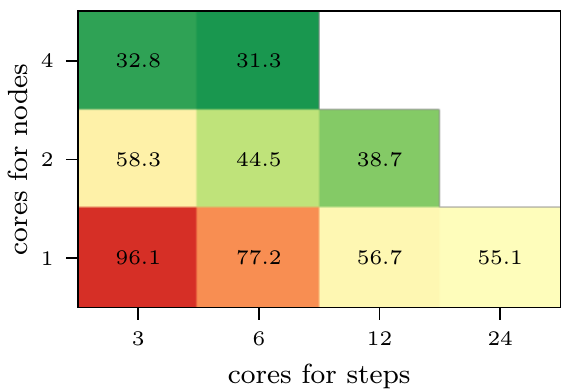}   
    \caption{Runtimes with different distribution of cores using PFASST-ER with $\matr{Q}$ for the Allen-Cahn equation.}
    \label{AC:time_Q}
\end{figure}
The two plots in Figure~\ref{AC:time_Q} show different combinations of cores used for step-parallelization ($x$-axis) and for node-parallelization ($y$-axis) with PFASST-ER and $\matr{Q}$.
Multiplying the numbers on both axes gives the total number of cores used for this simulation.
This is also the reason why there are two plots, because not all combinations are actually possible or meaningful.
Within each colored block the total runtime for this setup is given.
We can nicely see that using all available cores for parallelization across the step is by far not the most efficient way.
In turn, more than $4$ cores cannot be used for parallelization across the nodes, although this gives the best speedup.
Indeed, the best combination for this problem is to maximize node-parallelization first and then add step-parallelization ($31.3$ seconds with $4$ cores on the nodes and $6$ on the steps, lower picture).
This is about $1.8$ times faster than using $24$ cores for the steps alone and more than $5$ times faster than the serial PFASST-ER run.

Although using $\Qmat$ instead of $\QDmat$ in PFASST-ER is faster for this example, it is quite revealing to repeat the simulations using $\QDmat$.

  \begin{figure}[t] 
    \includegraphics[width=0.425\textwidth]{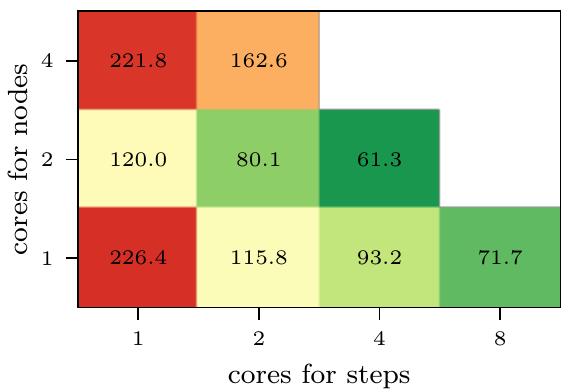}
    \includegraphics[width=0.425\textwidth]{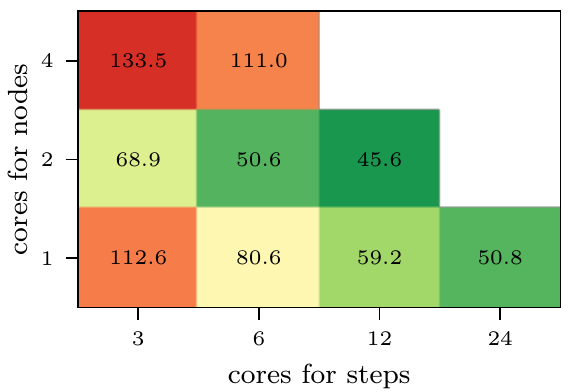}    
 \caption{Runtimes with different distribution of cores using PFASST-ER with $\QDmat$ for the Allen-Cahn equation.}
 \label{AC:time_T}
 \end{figure}

These results are shown in Figure~\ref{AC:time_T} and it is obvious that using as many cores as possible for the parallelization across the nodes now is not the optimal strategy.
Here, using $2$ cores on the nodes and $12$ on the steps is the most efficient combination, albeit still significantly slower than using PFASST-ER with $\Qmat$, even with the same combination.
The reason for this potentially surprising result is that solving the innermost linear systems heavily depends on the structure of these systems, in particular when using an iterative solver like GMRES.
Moreover, initial guesses are a crucial factor, too. 
For PFASST-ER, we use the current solution at node zero of the respective time-step as initial guess.
This is particularly suitable for the closest first nodes, but potentially less for later ones.
While both effects did not lead to significant variations in the time spent for solving the linear systems when using $\Qmat$, it does produce a severe load imbalance when using $\QDmat$.
More specifically, using $4$ cores for the nodes and only $1$ for the time-steps, i.e.\ exploiting only parallelization across the nodes, the first core takes about $118.2$ seconds for all linear system solves together at the first node, while the last core takes about $194.6$ seconds on the last node.
Therefore, using $2$ cores on the nodes, where core $1$ deals with nodes $1$ and $3$ and core $2$ with $2$ and $4$ is the ideal choice.
This is precisely what has been done for Figure~\ref{AC:time_T}, leading to the best speedup with $2$ cores on the nodes.

\bigskip
 
\begin{figure}[t]
\includegraphics[scale=0.9]{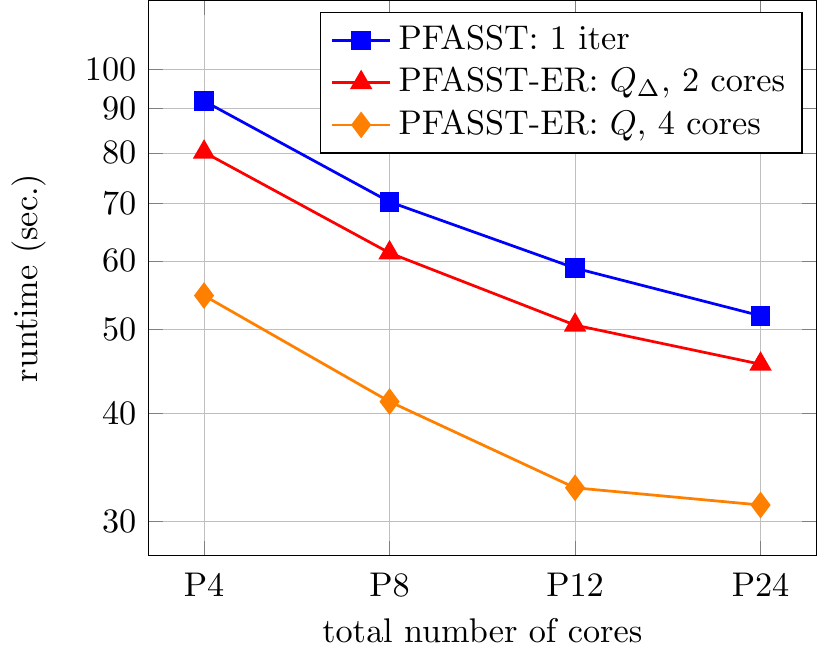}
  \caption{Runtimes for the three best variants, Allen-Cahn example.}
  \label{AC:comparison}
\end{figure}

In Figure~\ref{AC:comparison} we now summarize the best results: PFASST with one inner Newton iteration in comparison to PFASST-ER using $\QDmat$ and $2$ cores on the nodes and PFASST-ER using $\Qmat$ with $4$ cores on the node.
The plot shows the simulation time for each variant based on the number of processors used in total.
We see that PFASST-ER is always much more time efficient in doing the calculations than PFASST, with another significant gain when using $\Qmat$ instead of $\QDmat$.
Now, since PFASST-ER adds another direction of parallelization compared to PFASST, we can not only increase parallel efficiency as shown, but also extend the number of usable cores to obtain a better time-to-solution (but not necessarily a better parallel efficiency).
\begin{figure}[t]
    \includegraphics[scale=0.9]{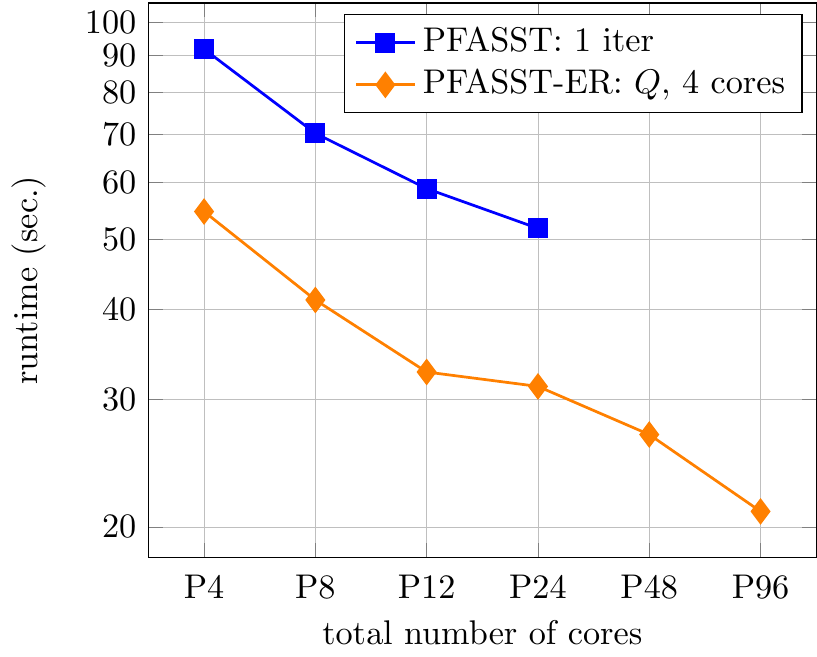}
    \caption{Runtimes for differnt number of processors, Allen-Cahn example.}
    \label{AC:comparison2}
\end{figure} 
This has been done in Figure~\ref{AC:comparison2}: taking $48$ or $96$ cores in total further reduces the computing time for $24$ time-steps. 
With PFASST-ER, the number of resources that can calculate parallel-in-time is no more limited by the number of time-steps, but can be increased by the factor given by the number of quadrature nodes.

\subsection{Gray-Scott equations}

The second example we present here is the Gray-Scott system \citep{pearson1993complex}, which is given by
\begin{align*}
u_t &= D_u \Delta u -2uv + F(1-u),\\
v_t &= D_v \Delta v +2uv - (F+K)v,
\end{align*}
on the spatial domain $[0,1]\times[0,1]$.
As initial condition we choose a circle with radius $0.05$ centred in the spatial domain, where $u=0.5$ and $v=0.25$ at the inside and $u=1.0$ and $v=0$ outside of this circle.  
We use $D_u=10^{-4}$, $D_v=10^{-5}$ and set a feed rate of $F=0.0367$ and a kill rate of $K=0.0649$.
This leads after some time to a process similar to cellular division and is known as ``mitosis''.
We refine the spatial domain with $128$ points in each dimension on the fine level and with $64$ on the coarse one, using standard finite differences.
We discretize every time-step of size $\Delta t =1$ with $4$ quadrature nodes and run the simulation again for $24$ time-steps.
The results will be presented very similar to the ones for the Allen-Cahn equation in the previous section.
We will omit the case of PFASST with more than one inner Newton iteration, though.

 \begin{figure}[t]
   \includegraphics[scale=0.9]{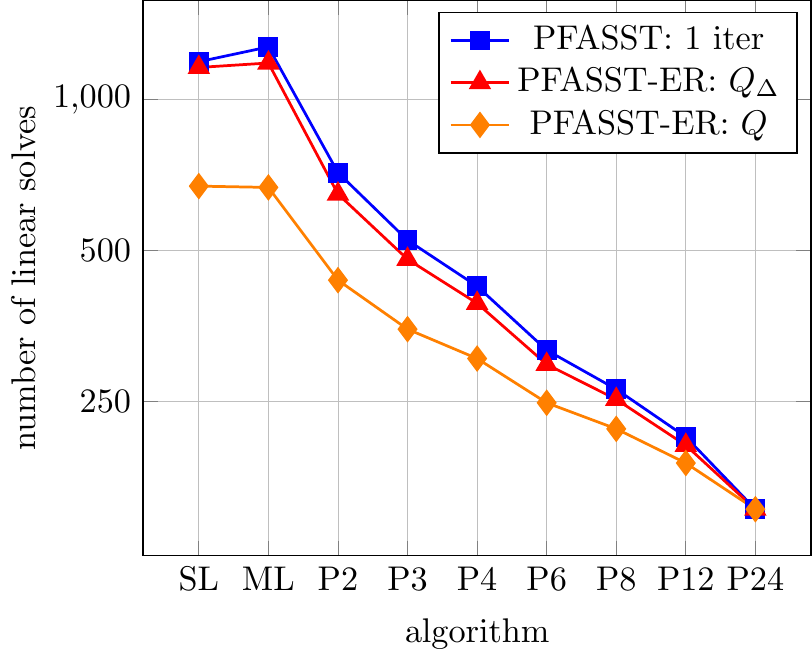}
     \caption{Number of linear solves for the Gray-Scott example, all methods run serial on the nodes.}
     \label{GS:iterations}
 \end{figure}

We start again looking at the total number of linear solves the different algorithms need to perform.
Figure \ref{GS:iterations} shows the number of linear solves for the methods, which run until a residual tolerance of $10^{-12}$ is reached.
The results look quite similar to the ones we got for the previous example, with one critical difference: The difference between the $\Qmat$-variant of PFASST-ER and the other algorithms becomes smaller more rapidly the more parallel time-steps are used.
In particular, it needs about the same number of inner solves as the others for $24$ cores.
Thus, one can expect that the runtime will increase when using PFASST-ER with $\Qmat$, while it stayed about the same in the case of the Allen-Cahn example.
\begin{figure}[t]
    \includegraphics[scale=0.9]{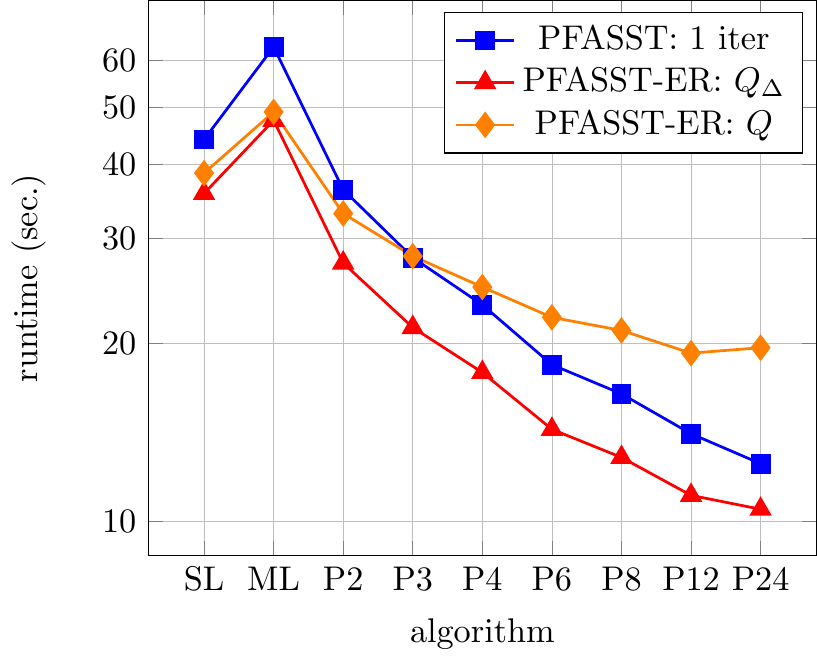}
    \caption{Time to solution for Gray-Scott with parallelization only across time-steps.}
    \label{GS:timing}
\end{figure}
This is precisely what we can see in Figure~\ref{GS:timing}.
The more parallel time-steps are run, the less efficient PFASST-ER with $\Qmat$ in this variant becomes. 
Already at $3$ parallel steps, it is as costly as the original PFASST version, at least when parallelization across the nodes is not considered.

Now, adding node-parallelization, the findings are again similar to the ones in the previous section:
\begin{figure}[t] 
   \includegraphics[width=0.425\textwidth]{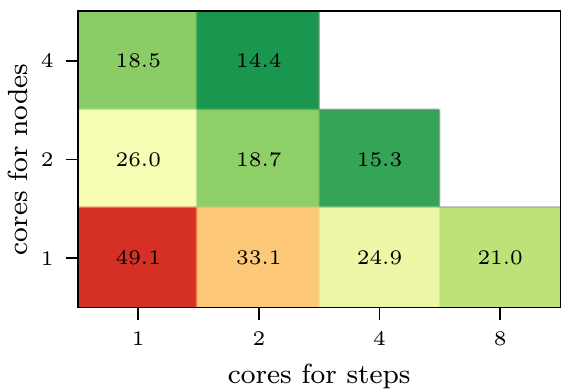}   
   \includegraphics[width=0.425\textwidth]{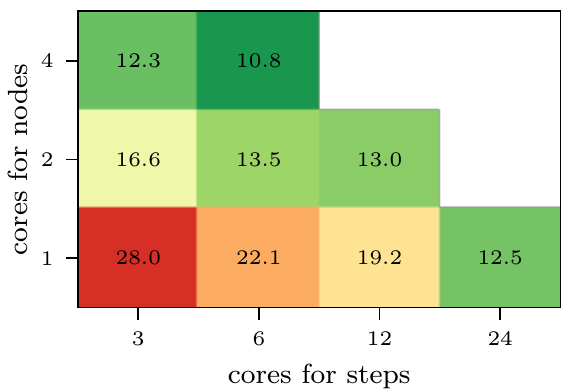}   
   \caption{Runtimes with different distribution of cores using PFASST-ER with $\matr{Q}$ for the Gray-Scott equations.}
      \label{GS_timeQ}
\end{figure}
Figure~\ref{GS_timeQ} shows that PFASST-ER with $\Qmat$ is still more efficient than using PFASST.
In particular, using more cores on the nodes is better and the best combination is again $4$ cores on the nodes and $6$ on the steps.
\begin{figure}[t] 
    \includegraphics[width=0.425\textwidth]{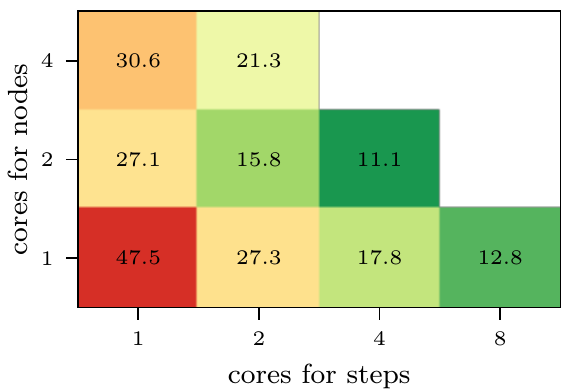}
    \includegraphics[width=0.425\textwidth]{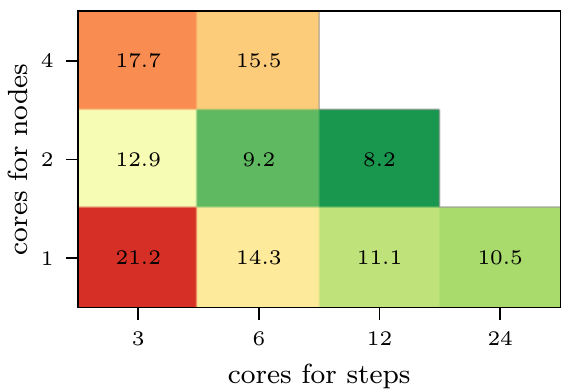}    
    \caption{Runtimes with different distribution of cores using PFASST-ER with $\QDmat$ for the Gray-Scott equations.}
    \label{GS_timeQD}
\end{figure}
Again, this changes when considering PFASST-ER with $\QDmat$ as in Figure~\ref{GS_timeQD}, where the ideal setup uses only $2$ cores on the nodes, but $12$ on the steps.
This is again due to load imbalances of the innermost linear solves.
However, note the key difference to the previous results: The fastest run of the $\QDmat$-variant is now faster than the one of the $\Qmat$-variant.

\begin{figure}[ht]
    \includegraphics[scale=0.9]{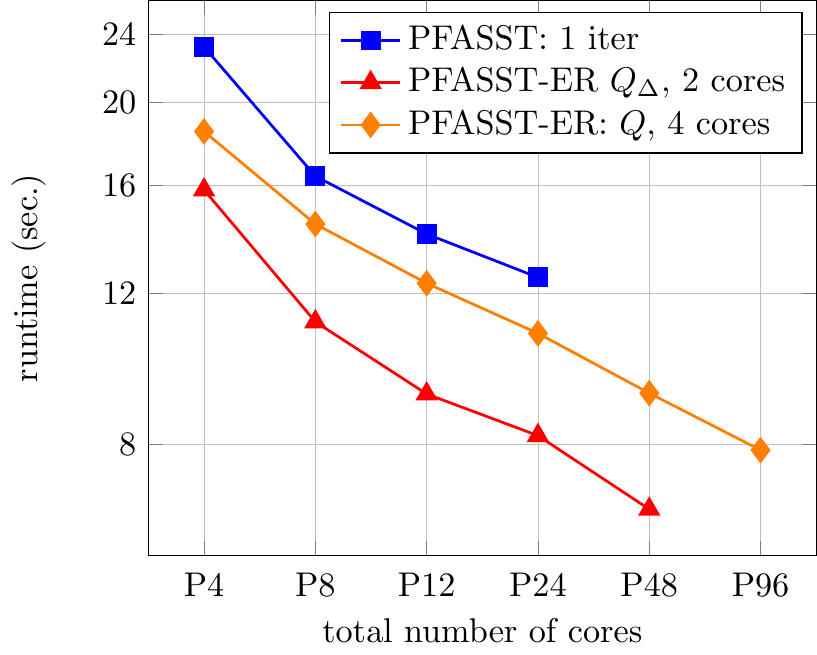}
    \caption{Runtimes for the three best variants, Gray-Scott example.}
    \label{GS:allT}
\end{figure}  

In Figure~\ref{GS:allT} we now give an overview about the best results: 
If we use parallelism across the nodes in a suitable way, both PFASST-ER versions are more efficient based on the simulation time than the classical PFASST algorithm.
Both can be used to extend the scaling capabilities beyond the number of time-steps, and both scale rather well in this regime.
Note, however, that the $\QDmat$-variant can here only leverage $2\times 24$ cores.
It is then faster than the $\Qmat$-variant with twice as many cores.

\section{Conclusion and outlook}

Nowadays supercomputers are designed with an ever increasing number of processors. 
Therefore we need our software and the underlying numerical algorithms to handle this increasing degree of parallelism.
Time-parallel integrators are one promising research direction, with quite a number of different approaches. 
Some approaches parallelize each individual time-step and others act on multiple time-steps simultaneously.
In this paper we have introduced a solver that works parallel across the method as well as parallel across the steps.
More precisely, we could combine node-parallel spectral deferred corrections with the parallel full approximation scheme in space and time.
While PFASST allow to compute multiple time-steps simultaneously and target large-scale parallelism in time, the new version called PFASST-ER presented here extend this idea with a very efficient small-scale parallelization for every single time-step itself. 
The scaling studies showed that a combination of both concepts seems to be the most efficient way to solve time-dependent PDEs.
Here we tested two different preconditioners: ones using the traditional, triangular quadrature matrix $\QDmat$ and one using the original matrix $\Qmat$.
Both could be diagonalized and used as parallel-across-the-node preconditioners.
For the $\QDmat$-preconditioner, we saw load imbalances when using an inner iterative linear solver, but by grouping nodes we still could speed up the simulation beyond the number of parallel time-steps.
For the $\Qmat$-preconditioner, the overall number of iterations was lower and time-to-solution was faster.
Adding node-parallelization, parallel efficiency could be increased and speedup extended when compared to PFASST.
Both PFASST-ER versions lead in the end to better scaling than the classical PFASST algorithm.

During our experiments we saw that it is not clear a priori, which combination of node- and step-parallelization is the most efficient one.
This leads to a lot of, potentially irrelevant runs to find the sweet spot.
Here, a performance model and a suitable convergence theory are needed to at least narrow down the relevant options.
This has to be accompanied by more numerical tests, relating e.g.\ model parameters with load imbalances, to identify the limits of this approach.

\section*{Acknowledgements}
The authors thankfully acknowledge the financial support by the German Federal Ministry of Education and Research through the ParaPhase project within the framework ``IKT 2020 - Forschung f\"ur Innovationen'' (project number 01IH15005A).

\bibliographystyle{abbrvnat}
\bibliography{refs}

\end{document}